# Ultrafast supercontinuum spectroscopy of carrier multiplication and biexcitonic effects in excited states of PbS quantum dots

F. Gesuele[1*], M. Y. Sfeir[2], W.-K. Koh[3], C. B. Murray[3,4], T. F. Heinz[5], and C. W. Wong[1*]

[1]*Optical Nanostructures Laboratory, Center for Integrated Science and Engineering, Solid-State Science and Engineering, and Mechanical Engineering, Columbia University, New York, NY 10027*
[2]*Center for Functional Nanomaterials, Brookhaven National Laboratory, Upton, NY 11973*
[3]*Department of Chemistry, University of Pennsylvania, Philadelphia, PA 19104*
[4]*Department of Materials Science and Engineering, University of Pennsylvania, Philadelphia, PA 19104*
[5]*Departments of Physics and Electrical Engineering, Columbia University, New York, NY 10027*

**Abstract**

We examine the multiple exciton population dynamics in PbS quantum dots by ultrafast spectrally-resolved supercontinuum transient absorption (SC-TA). We simultaneously probe the first three excitonic transitions over a broad spectral range. Transient spectra show the presence of first order bleach of absorption for the $1S_h$-$1S_e$ transition and second order bleach along with photoinduced absorption band for $1P_h$-$1P_e$ transition. We also report evidence of the one-photon forbidden $1S_{h,e}$-$1P_{h,e}$ transition. We examine signatures of carrier multiplication (multiexcitons for the single absorbed photon) from analysis of the first and second order bleaches, in the limit of low absorbed photon numbers ($\langle N_{abs} \rangle \sim 10^{-2}$), at pump energies from two to four times the semiconductor band gap. The multiexciton generation efficiency is discussed both in terms of a broadband global fit and the ratio between early- to long-time transient absorption signals.. Analysis of population dynamics shows that the bleach peak due to the biexciton population is red-shifted respect the single exciton one, indicating a positive binding energy.

[*]*corresponding authors: fg2251@columbia.edu, cww2104@columbia.edu*

There is tremendous interest in the properties of quantum dot (QD) based-materials for third-generation photovoltaics [1, 2]. The strong spatial confinement of electronic wave function leads to an enhancement of the Coulomb carrier-carrier interaction. This effect together with momentum relaxation and the wide separation of the energy levels (limiting phonon-assisted transition) can produce multiple excitons efficiently. This effect for single photon absorption is named carrier multiplication (CM) and can be explained [3, 4] in terms of an increased impact ionization rate, where a high energetic absorbed photon produces a "hot exciton" which decays, leaving energy to the valence band to generate an additional exciton. In addition to spatially-separated multiple excitons [5], this effect is a promising mechanism to enhance the solar cell quantum efficiency, by increased photocurrent [6] above the Shockley-Queisser limit [7].

In recent years evidence of carrier multiplication has been reported but the results seem to depend strongly on sample preparation, experimental conditions and interpretation of the results [8-16]. There has been debate on the real efficiency in view of the practical applications, although recent studies [17, 18] in chemically synthetized materials suggest the evidence of increased photocurrent due to carrier multiplication. Among other pump-probe spectroscopic techniques such as time-resolved photoluminescence, intraband absorption, second harmonic, sum-frequency or terahertz generation [19-22], ultrafast interband transient absorption technique [23, 24] has been used to investigate the carrier population dynamics by measuring the pump induced change in absorption. Results reported in literature almost all refer to the lowest energy $1S_h$-$1S_e$ absorption feature. So far the results do not take into account the homogeneous and ensemble broadening of first exciton peak and its dispersive behavior, while it has been clearly shown that the few or single isolated near-infrared lead chalcogenide nanocrystals have a spectral broadening of ~ 100 meV at room temperature [25, 26]. The broad-band spectral dispersion has not been rigorously examined in carrier multiplication studies and its related efficiencies, though recent studies [10, 12, 27, 28] have examined the bleach dynamics at discrete wavelengths within only the first exciton transition.

In this Letter we investigate the multiexciton generation and dynamics in PbS QDs with a supercontinuum ultrafast transient absorption (SC-TA) technique, encompassing the first three excitonic interband transitions simultaneously ($1S_h$-$1S_e$, $1P_h$-$1P_e$ and $1S_{h,e}$-$1P_{h,e}$). In our measurements we consider the excitation of the QDs at pump energies from two to four times the semiconductor band gap while probing the absorption changes with a femtosecond white-light supercontinuum probe. The simultaneous observations of the spectrally-resolved ultrafast dynamics enable a global analysis of the carrier dynamics.

Sample investigated in this work are oleic-capped PbS QDs suspended in trichloroethylene (TCE) solvent and synthesized in an oxygen-free glove box environment [29]. After fresh preparation the samples were rapidly transferred, inside the glove box, to fused quartz cuvettes and sealed in order to prevent oxygen exposure during measurements. The optical absorption spectra of investigated samples, reported in Figure 1a, are measured in the spectral range of supercontinuum probe (800 to 1600 nm). The narrow size distribution of QD samples allows the clear identification of the set of optical absorption peaks. Our measurements evidence the first and second absorption peaks in spectral position reported in literature [30, 31]. Some samples also show the presence of a weak absorption peak in-between. While the first exciton peak is traditionally associated to the $1S_h$-$1S_e$ transition, the identification of the second peak in linear absorption in PbS QDs has been the subject of debate recently [30, 32-34].

Following the energy-level calculation presented in literature and based on the four-band $\mathbf{k}\cdot\mathbf{p}$ envelope model [35] with account of band anisotropy [36] or wave function anisotropy [30] we attribute the second peak to the one-photon allowed $1P_h$-$1P_e$ transition [37]. The minima in-between is associated to the one-photon absorption (1PA) forbidden $1S_{h,e}$-$1P_{h,e}$ transitions [33]. In particular the isotropic model cannot explain the 1PA transition among the S and P levels because of the wavefunction parity, without inclusion of the energy band anisotropy. Recent work [30] pointed that symmetry breaking in the wave function can explain the 1PA peaks at the $1S_{h,e}$-$1P_{h,e}$ transitions as well as two-photon absorption peaks at the $1S_h$-$1S_e$ and $1P_h$-$1P_e$ transitions.

The experimental setup consists of a 1 mJ Ti:Sapphire 1 kHz femtosecond regenerative amplifier combined with an optical parametric amplifier (OPA) that allows generation of sub-100 fs pulses in the ultraviolet, visible, and infrared spectrums (Newport Spectra-Physics). This system, as illustrated in Figure 1b, is coupled to a supercontinuum transient absorption spectrometer with approximately 100 fs temporal resolution in a window up to 3 ns. The detection consists of a pair of high-resolution (512 pixel) multichannel detector array referenced at a high-speed data acquisition system (Ultrafast Systems). In this technique the samples are optically pumped using the spectrally tunable (240 to 2600 nm) femtosecond pulses generated in the OPA and probed for transmission changes using the generated white-light supercontinuum.

We observe that average probe power is few hundreds of nW over a broad range (850-1600nm), and it's definitely less that the pump power even for low-power intensity measurements. So we assume that the probe only create a small perturbation, while the exciton dinamics is fully determined by the pump. The samples are vigorously stirred in the trichloroethylene solution during measurements to avoid charge-multiplication-like artifacts such as fast decay signatures due to photoionization leading to QD charging [38]. Moreover, in order to avoid artifacts due to multiphoton absorption, the number of absorbed photons $\langle N_{abs} \rangle$ is estimated as the product of the pulse fluence measured in photons per second and centimeter squared ($j_p$) and the quantum dot absorption cross section at the pump wavelength ($\sigma_a$): $\langle N_{abs} \rangle = j_p\ \sigma_a$ [39, 31, 40].

In Figure 2(a, b, c) we illustrate a typical spectrally resolved differential absorption for QD samples with first exciton absorption centered at 1310 nm. The two-dimensional time-resolved spectrogram covers the energy of the first three excitonic transitions and up to a temporal range of 1 ns starting from pump probe coincidence. Time and spectral resolved measurements are corrected by the probe chirp which is independently measured in the same experimental condition by using solvent in place of nanocrystal solution.

Measurements were performed with excitation pump energies 2, 3 and 4 times the gap energy (HOMO-LUMO transition) for the same estimated number of absorbed photons (~0.2). From the 2D time- and spectral-resolved absorption, simultaneous bleaches corresponding to the first and second transitions can be distinguished as more pronounced at early time.

Figure 2d is an example differential absorption trace at a fixed time delay of 1.5 ps. We clearly distinguish two bands of bleaches. The first one corresponds to the fundamental excitonic peaks measured in linear absorption. In-between these two bands there is a photo-induced absorption band and, corresponding to the minimum of linear absorption, we can observe a region with weak differential absorption, even with high pump fluences. We ascribe this behavior to the 1PA forbidden $1S_{h,e}$-$1P_{h,e}$ transition. The photoinduced absorption region together with the second order bleach band can be ascribed to the higher order transition $1P_h$-$1P_e$. Explanation of this effect lies in a red-shift of this transition due to the presence of the $1S_h$-$1S_e$ exciton created by the pump pulse and its interaction with the exciton due to probe pulse [31, 38]. The positive band presents an asymmetric non-Gaussian lineshape probably indicating the superposition of a pure electron and pure hole intraband transitions or excited state absorption effect. In Figure 2e we plot one example of the (normalized) time evolutions of the $1S_h$-$1S_e$, $1P_h$-$1P_e$ bleaches and maxima of photoinduced absorption band, for $2E_g$ and with $\langle N_{abs} \rangle$ ~1.3. In this regime, no CM is expected but the presence of multiexcitons due to multiphoton absorption is possible. With respect to the first order ($1S_h$-$1S_e$) bleach, we observe that the second order ($1P_h$-$1P_e$) bleach has a stronger and faster decay component and a minimum at earlier time. This can be understood considering that, in the second order bleach, the relaxation processes from the higher order transition happen at hundreds of femtoseconds time-scale. If we compare the first and second transition bleaches we can easily distinguish three phases: assuming instantaneous absorption respect to the time resolution of our setup, the initial rise of second transition bleach to its maximum correspond to relaxation from higher order states to $1P_h$-$1P_e$. The interval between the maximum of the first and second bleaches refers to relaxation from $1P_h$-$1P_e$ to $1S_h$-$1S_e$ states, and the long range exponential decay with comparable lifetime in each of the three example cases indicates that the non-radiative Auger recombination process involves not only the first but as well the second bleach. This is reasonable considering that the exciton generation and recombination is a process that involves each of these bands because it changes simultaneously the electron and hole occupation numbers. The Auger recombination of one exciton occurs, leaving energy for electron and/or hole to move to a higher state, from which it will relax through the intermediate excited states. In Figure 2f, we illustrate the dispersive bleach for wavelengths around the first transition. The spectral variation of initial rise and photoinduced absorption can be observed for early-time dynamics, corresponding to previous observations [10]. The longer-wavelength positive band has also been suggested to arise from excited state absorption [47]

Estimation of the multiple exciton generation process is based on the observation [23] of the Auger recombination of a biexciton in the fundamental state $1S_h$-$1S_e$ into a single high energetic exciton. As shown in Figure 3, this process happens in tens of picoseconds and, from the ratio of the relative maximum at early time (*a*-*b*) to the long-time asymptotic value (*b*), the multiexciton efficiency can be extracted. To avoid distortion of the multiexcitons generated by multi-photon absorption, our SC-TA measurements are performed with low pump fluence, in the limit of low absorbed photon number ($\langle N_{abs} \rangle$ ~ 0.01).

In Figure 3 we illustrate example series of supercontinuum bleaches for both the $1S_h$-$1S_e$ (panel a,c) and $1P_h$-$1P_e$ (panel b,d) bands for different pump fluences and pump energies $2E_g$ (Figure 3a and 3b) and $4E_g$ (Figure 3c and 3d). The inset in Figure 3a and 3c summarizes the multiexciton efficiency [*(a-b)/b=a/b-1=R-1*] as a function of $\langle N_{abs} \rangle$, for the first transition bleach. We assume a poissonian distribution of absorbed photons and we have fitted the ratio of transient absorption at short and long delays in fig 3(a,c) with [16]

$$R = \frac{J\sigma_a \delta \cdot QY}{1 - \exp(-J\sigma_a)}$$

The asymptotic behavior at low pump fluence ($\langle N_{abs} \rangle$ ~0.01) at $4E_g$ is the signature of CM presence. At the same time the zero asymptotic value for the $2E_g$ plot evidence that multiexcitons can be generated only by multiphoton absorption and no significant CM is observed. Figures 3b and 3d are the corresponding pump-dependent bleaches for the $1P_h$-$1P_e$ transition at $2E_g$ and $4E_g$, where we illustrate the maximum at early time (*c*-*d*) and the long-time asymptotic value (*d*) for the $1P_h$-$1P_e$ transition. We observe that, in analogy to the $1S_h$-$1S_e$ dynamics, at low $\langle N_{abs} \rangle$ values the $1P_h$-$1P_e$ ratio $(c\text{-}d)/d$ at $4E_g$ asymptotes to a non-zero value highlighting the presence of carrier multiplication while at $2E_g$ asymptotes to zero indicating the absence of carrier multiplication. This higher-order transition dynamics serves as an alternative approach to determine the carrier multiplication. In absence of multiexctions the ratio tends to zero because the change in bleach is constant over initial picosecond timescales.

To account for the spectral dispersion we evaluate the supercontinuum ultrafast dynamics with global fits via singular value decomposition. In the exponential regression analysis we impose that each different wavelength is composed of the same three lifetimes but with different weights. This corresponds to the estimation of the overall 2D measured differential absorption with:

$$\Delta\alpha L(\lambda,t) = a_1(\lambda) exp\left(-\frac{t-t_0}{t_1}\right) + a_{Auger}(\lambda) exp\left(-\frac{t-t_0}{t_{Auger}}\right) + a_\infty(\lambda) exp\left(-\frac{t-t_0}{t_\infty}\right)$$

In our model we impose three lifetimes: the first ($t_1$) at hundreds of femtoseconds which correspond to relaxation from higher order transition; the second ($t_{Auger}$) at tens of picoseconds which correspond to the non-radiative Auger recombination process; and the third ($t_\infty$) at hundreds of nanoseconds which correspond to single exciton radiative lifetime [41]. The coefficients of the fit represent the molar differential absorption of the species multiplied by their concentration in the solution. So far $a_{Auger}$ and $a_\infty$ corresponds respectively to biexciton and single exciton species [42, 43]. In this way, in the limit of low $\langle N_{abs} \rangle$, the ratio $a_{Auger}/a_{\infty}$ represents the spectrally resolved CM efficiency.

In Figure 4 we plot the coefficients of our exponential regression for $4E_g$ and with $\langle N_{abs} \rangle$ at 0.04, with the obtained CM efficiency (black dotted) in panel b. We observe that it is not constant over the width of first $1S_h$-$1S_e$ band transition (superimposed as red line) but increase monotonically with wavelength. This can be understood considering that the minima for the coefficients $a_{Auger}$ and $a_{\infty}$ happen at different wavelengths with the $(a_{Auger})_{min}$ red-shifted; thus indicating that the early-time exciton population, which decays with $a_{Auger}$ lifetime, is red-shifted with respect to late time spectra. We observe that from the global fitting procedure, the coefficient at 558fs present positive value for SS transition (indicating the increase of bleach), while negative for PP (indicationg decrease of bleach). This can be linked to the decay from the PP state to SS state.

The physical explanation of this effect lies in the population of $1S_h$-$1S_e$ state. In Figure 5a, we compare the differential spectra (at $4E_g$ in presence of CM) at early time maximum (~ 2.5 ps) with one at late time (~ 650 ps) where one can observe a different lineshape and maxima. The early time spectrum (maximum bleach) is red-shifted with respect to the late time spectrum (after Auger recombination) by ~ 7meV (a convolution of single exciton and biexciton population bleaches) for this particular example. This happens for $4E_g$ as well $2E_g$ with high $\langle N_{abs} \rangle$ as shown in Figure 5b. The presence of a second exciton in the nanocrystal introduces an attraction or repulsion (positive or negative) binding energy [44]. In these symmetric core nanocrystals the Coulomb interaction tends to spatially distribute the charge in order to obtain a net positive binding energy [45, 46]. Hence the presence of second exciton results in a biexciton energy less than twice the single-exciton energy which corresponds to a red-shifted minimum of bleach. To our knowledge this is the first observation of the biexciton effect and positive biexciton binding energy in PbS nanocrystals supercontinuum transient absorption analysis.

In conclusion we have studied the ultrafast dynamics of PbS QDs with supercontinuum spectrally-resolved interband transient absorption. We have shown that the differential spectra present photo-induced absorption features for the 1PA forbidden $1S_{h,e}$-$1P_{h,e}$ transition. The presence of carrier multiplication effect can be observed in the higher ($1P_h$-$1P_e$) order bleaches in a similar way to the $1S_h$-$1S_e$ bleach. Moreover the dispersive effect must be taken into account in the carrier multiplication estimates due to a CM efficiency that increases with wavelength across the $1S_h$-$1S_e$ transition. This behavior has been explained with the red-shift of early time population due to the biexciton binding energy.

The authors acknowledge discussions with Jonathan Schuller, Jonathan Owen, James Misewich, James Yardley, David Reichman, and Charles Black. This material is based upon work supported as part of the Center for Re-Defining Photovoltaic Efficiency Through Molecule Scale Control, an Energy Frontier Research Center funded by the U.S. Department of Energy, Office of Science, Office of Basic Energy Sciences under Award Number DE-SC0001085. CBM and WKK acknowledge financial support from the NSF through DMS-0935165 and partial support by the Nano/Bio Interface Center through the National Science Foundation (NSEC DMR08-32802). Research carried out in part at the Center for Functional Nanomaterials, Brookhaven National Laboratory, which is supported by the U.S. Department of Energy, Office of Basic Energy Sciences, under Contract No. DE-AC02-98CH10886.

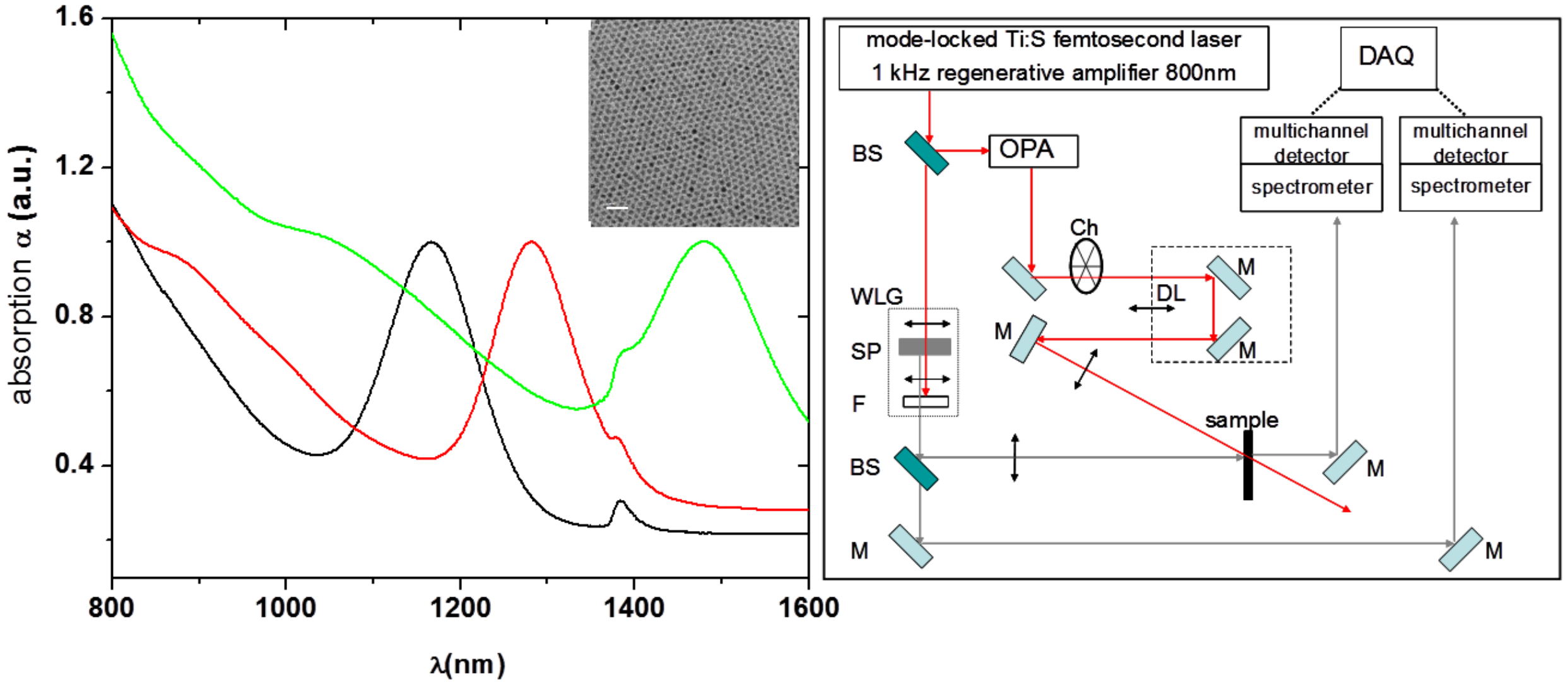

**Figure 1:** PbS nanocrystals for multiexciton generation and ultrafast dynamics studies. **(a)** Linear absorption spectra for varying sizes. Inset: high-resolution TEM of PbS nanocrystals. Scale bar: 10-nm. Absorption feature at~1400nm comes from the fused quartz cuvette used for measurements and gives no features in the time resolved data **(b)** Schematic of broadband ultrafast transient absorption double channel spectroscopic setup, with 800-nm mode-locked Ti:S femtosecond laser and multichannel high speed detector in the spectrometer. OPA: optical parametric amplifier; DL: delay line; M: mirror; BS: beam splitter; WLG: white light generator; SP: sapphire plate; DAQ: high-speed data acquisition; Ch: optical chopper; F: Filter.

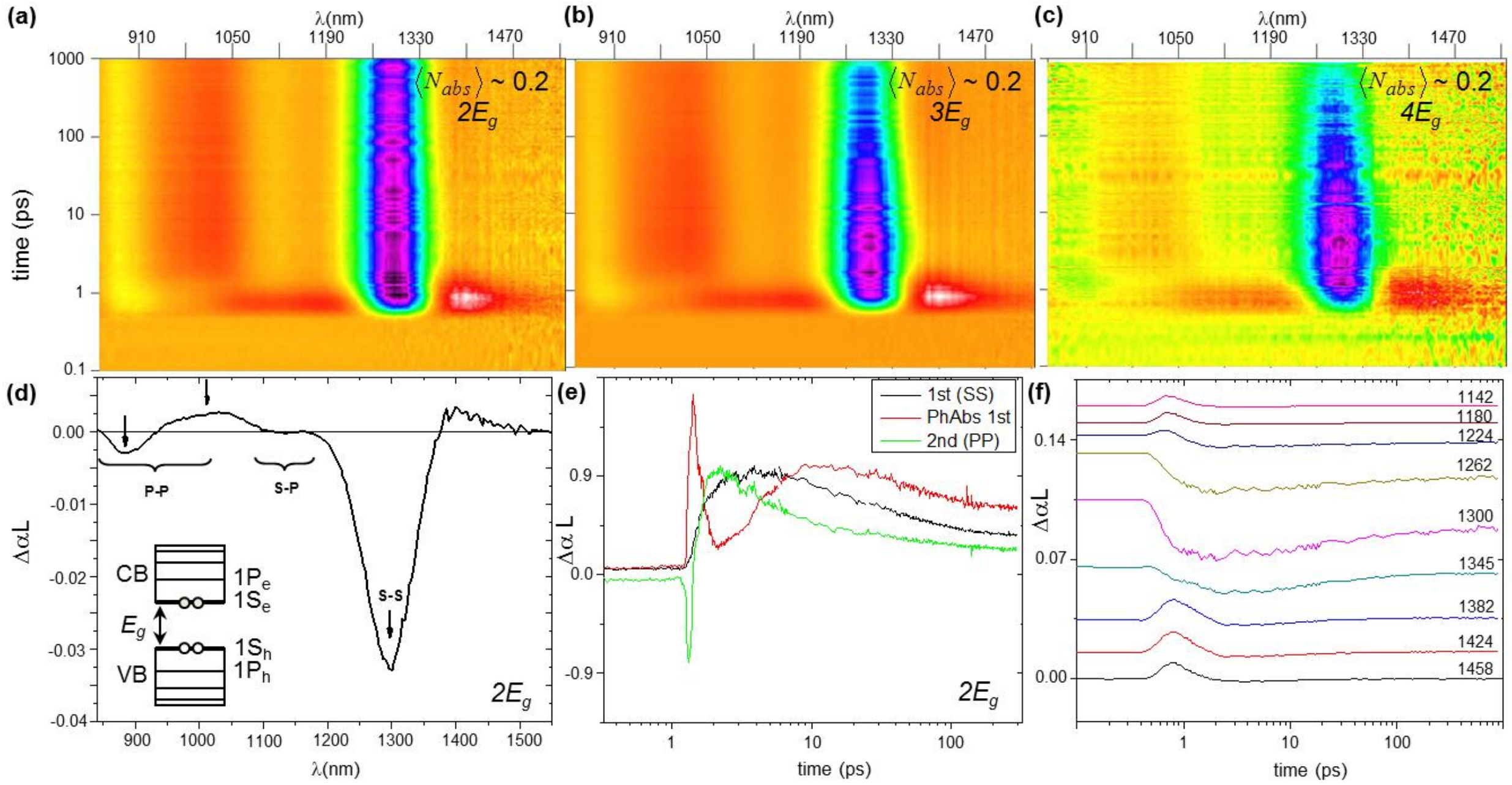

**Figure 2:** **(a)** to **(c)** Supercontinuum femtosecond transient absorption (SC-TA) for PbS nanocrystal with first exciton at 1310 nm and pumped at 640 nm ($2E_g$), 427 nm ($3E_g$) and 320 nm ($4E_g$) respectively. The pump fluence is attenuated to maintain $\langle N_{abs} \rangle$ ~ 0.2. **(d)** Differential spectrum at 1.5 ps delay pump-probe with $2E_g$ and $\langle N_{abs} \rangle$ ~ 1.3, corresponding at maximum bleach. Sample with first exciton at 1310 nm. We have evidenced the bleach for first and second transitions as well as a region of photoinduced absorption. In the region with small $\Delta\alpha$L we assign the 1PA forbidden SP transition. **(e)** Dynamics at first and second bleach (sign-inverted) and photoinduced absorption (arrows indicate corresponding wavelengths) band with the same conditions as panel **(d)**. Peaks are normalized and happen at different time delays. **(f)** Example bleach dynamics with dispersion around the first exciton peak (1310 nm), at $2E_g$ and high fluence ($\langle N_{abs} \rangle \sim 1$). Labels indicate the wavelength examined (in nm).

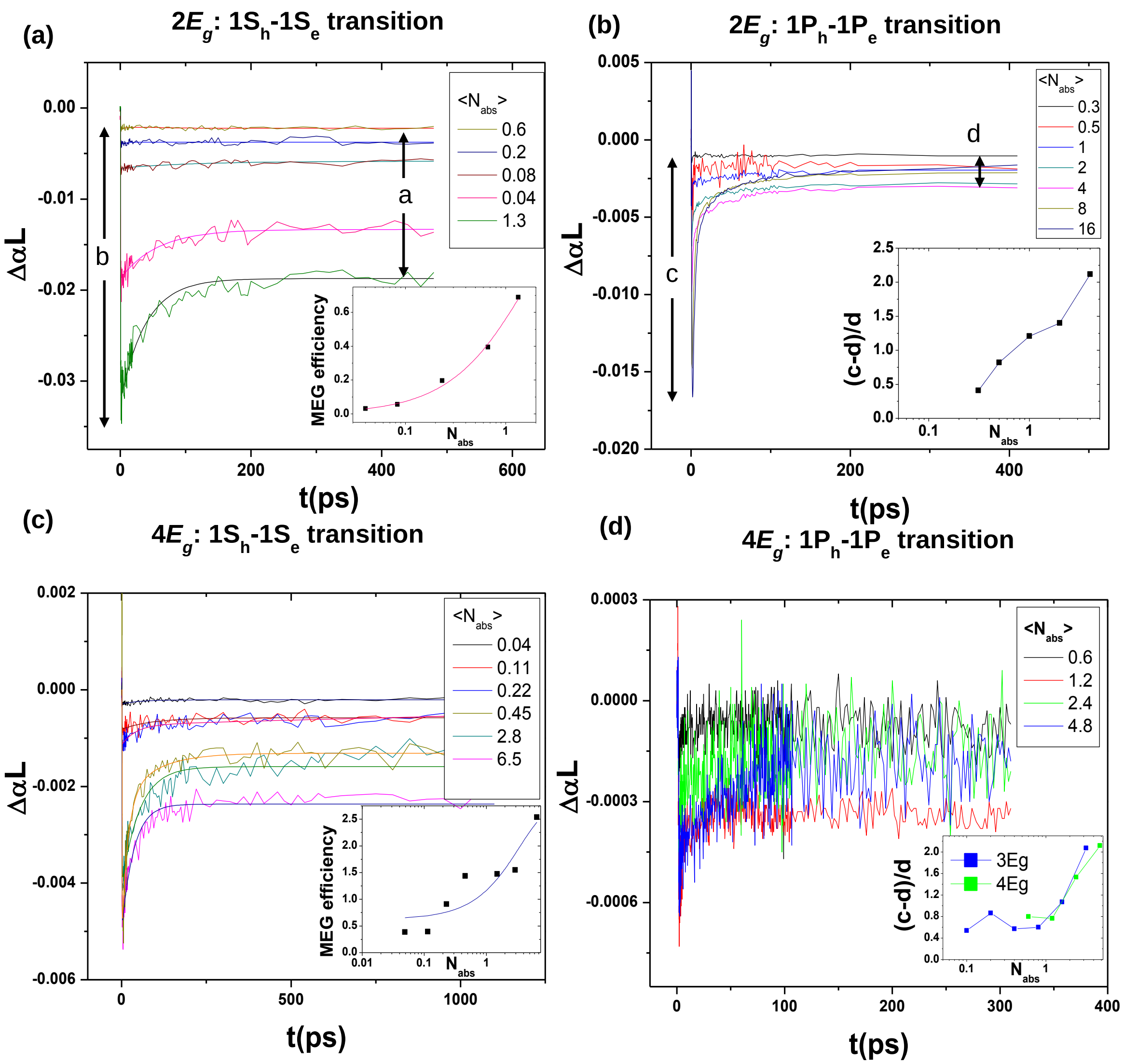

**Figure 3:** (**a** and **c**) Bleach dynamics for first excitonic transition for pump wavelengths at $2E_g$ and $4E_g$ respectively. Inset shows corresponding MEG efficiency, calculated from the ratio of early to late time which asymptotes to zero. The asymptotic value at low photon fluency indicates carrier multiplication presence. (**b** and **d**) Bleach dynamics for second excitonic transition for pump wavelengths at $2E_g$ and $4E_g$ respectively. Inset shows the figure of merit *(c-d)/d* for $2E_g$ (**b**), $3E_g$, and $4E_g$ (**d**).

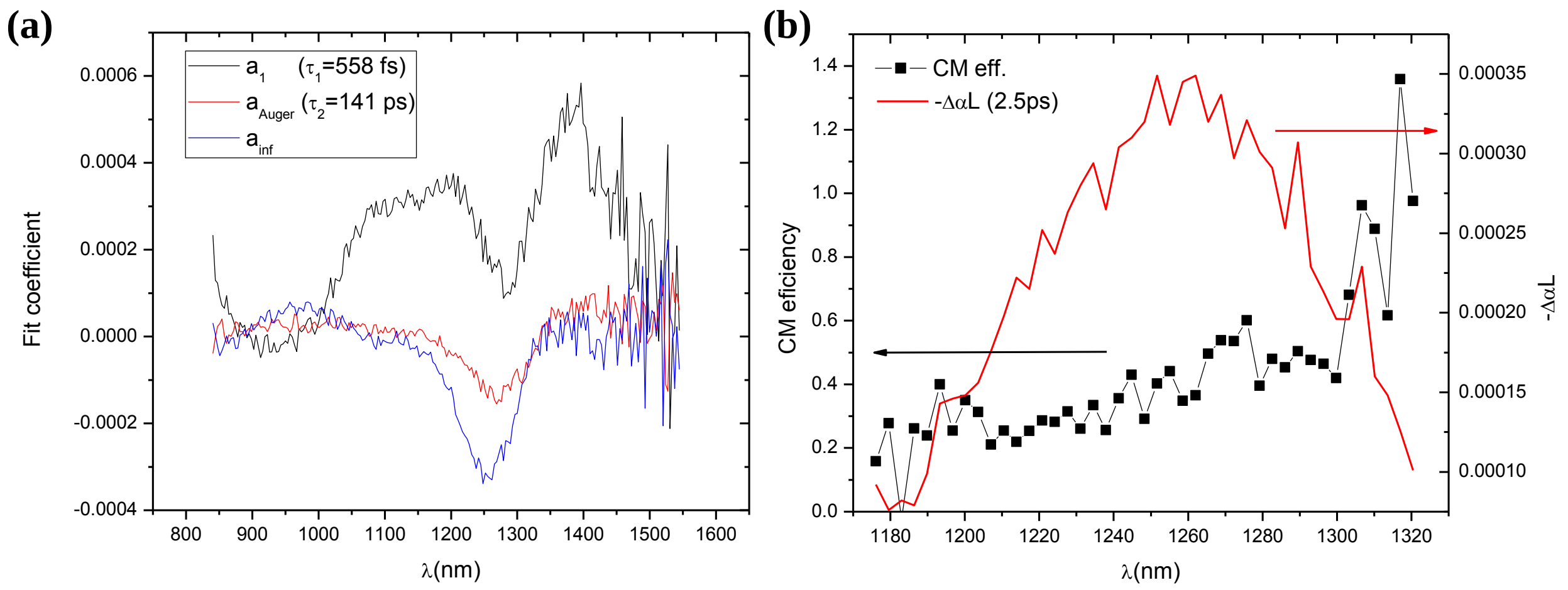

**Figure 4: (a)** Global fit coefficients for PbS nanocrystals pumped at $4E_g$ and $\langle N_{abs} \rangle \sim 0.04$. **(b)** The resultant spectrally-resolved CM efficiency (black dotted) obtained from the ratio between $a_{Auger}$ and $a_{\infty}$. The differential absorption spectrum is superimposed (red line) for comparison ease.

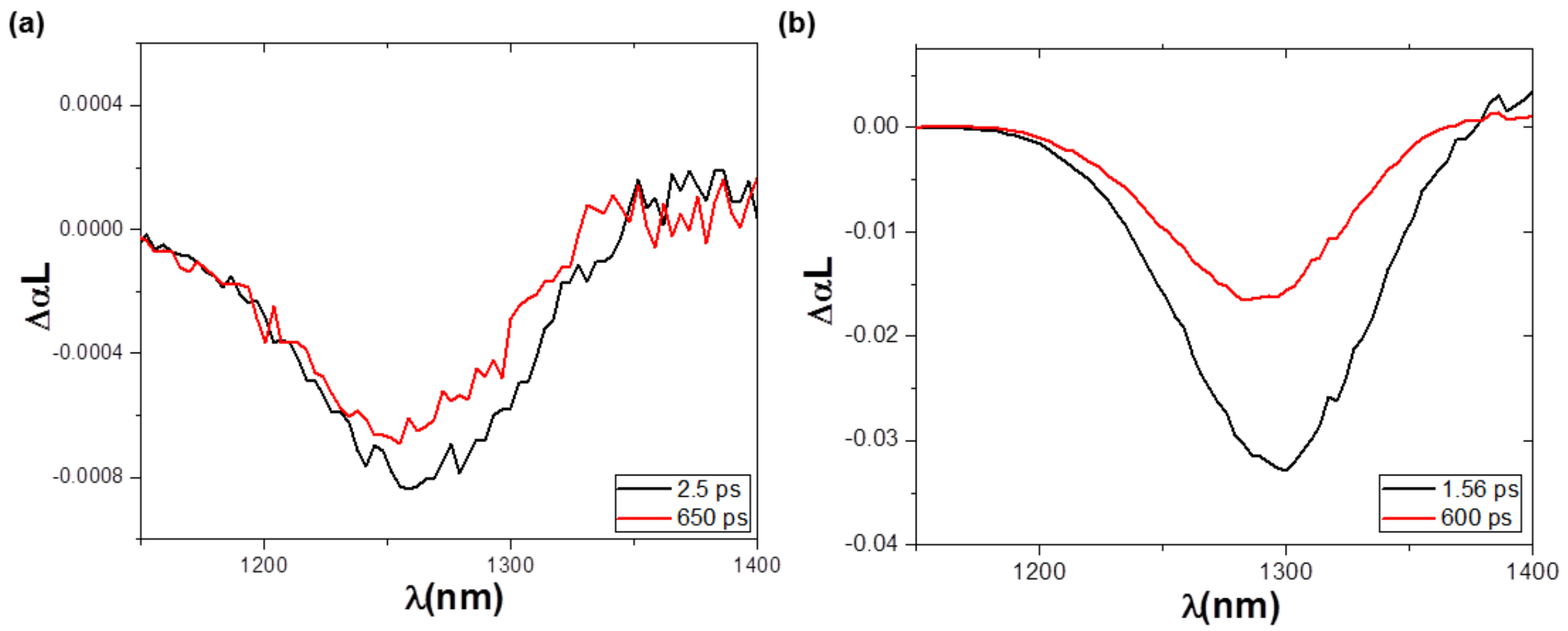

**Figure 5:** **(a)** Differential spectra for at early (2.5 ps) and late time (650 ps), showing an early red-shift due to biexciton effect. Pump at $4E_g$ and $\langle N_{abs} \rangle \sim 0.1$. **(b)** Same comparison for $2E_g$ and $\langle N_{abs} \rangle \sim 1.3$ shows the biexciton shift.